\documentclass[aps,showpacs,floatfix,twocolumn,pra,superscriptaddress]{revtex4}
\usepackage{amssymb,amsthm, amssymb, latexsym}
\usepackage{graphicx}
\usepackage{epstopdf}
\usepackage{amsmath}
\usepackage[english]{babel}
\usepackage{color}

\begin{document}

\title{Interaction-Induced Gradients Across a Confined Fermion Lattice}

\author{G. George Batrouni}
\affiliation{Universit\'e C$\hat o$te d'Azur, INLN, CNRS, France}
\author{Richard T. Scalettar}
\affiliation{
Department of Physics, University of California Davis, CA 95616, USA }

\begin{abstract}
  An imposed chemical potential gradient $A_\uparrow=d\mu_\uparrow/dx$
  on a {\it single} fermionic species (``spin up'') directly produces a
  gradient in the density $d\rho_\uparrow/dx$ across a lattice.  We
  study here the {\it induced} density inhomogeneity
  $d\rho_\downarrow/dx$ in the second fermionic species (``spin
  down'') which results from fermionic interactions $U$, even in the
  absence of a chemical potential gradient $A_\downarrow=0$ on that
  species.  The magnitude of $d\rho_\downarrow/dx$ acquired by the
  second species grows with $U$, while the magnitude of
  $d\rho_\uparrow/dx$ remains relatively constant, that is, set only
  by $A_\uparrow$.  For a given $A_\uparrow$, we find an interaction
  strength $U_*$ above which the two density gradients are equal in
  magnitude.  
We also evaluate the spin-spin correlations and
show that, as expected, antiferromagnetism is most dominant
at locations where the
local density is half-filled.  The spin polarization induced by
sufficiently large gradients, in combination with $U$, drives
ferromagnetic behavior.
In the case of repulsive interactions,
  $d\rho_\downarrow/dx = -d\rho_\uparrow/dx$.  A simple particle-hole
  transformation determines the related effect in the case of
  attractive interactions.
\end{abstract}

\maketitle

\section{Introduction}

Over the past several years, there has been a great deal of progress
in realizing antiferromagnetic correlations in trapped fermionic
gases\cite{hart15,duarte15,cheuk16,mazurenko17,brown16}. Therefore,
understanding the effects of chemical potential gradients has become a
central objective of quantum simulations of Hubbard Hamiltonians
modeling bosonic or fermionic atoms on optical lattices.  The most
common situation studied is a confining potential which rises smoothly
from the center of the system. An immediate consequence is the
presence and coexistence of a sequence of phases, superfluid and Mott
insulator in the bosonic
case\cite{batrouni02,demarco05,campbell06,barankov07,sun09}; and band
insulator, metallic, and antiferromagenetic insulator for
fermions\cite{jordens08,mathy12,fuchs11,schneider08}, as one moves
radially outward from the trap center.  With two fermionic species,
the focus is almost exclusively on the situation in which the
confining potential acts equally on ``spin up'' and ``spin down'', so
that the densities of the species, although varying spatially, do so
in lock step with each other.

In addition to the density variation across the lattice, and the
accompanying local phases, the behavior of the thermodynamics is of
considerable interest: The prospect of regions of the gas acting as
local entropy reservoirs has implications for adiabatic cooling as the
lattice is turned on, and the possibility of achieving longer range
antiferromagnetic correlations\cite{fuchs11,paiva11}

Although spin-dependent potentials are atypical in the condensed
matter context, they are possible to achieve in the case of optically
trapped atoms where the fermionic species correspond to different
hyperfine states. As an example, ``spin-dependent'' optical lattices
have been discussed some time ago and their usefulness in thermometry
and prospect for achieving exotic pairing states
explored\cite{mckay10,mandel03,feiguin09,liu04}.  Similarly, the
properties of disordered systems in which the {\it randomness} is spin
dependent (and the interactions are attractive) have been explored
within a Bogliubov-de Gennes approach leading to unusual gapless
superconducting phases and phase transitions
\cite{nanguneri12,jiang13}.  In the condensed matter context, such
spin dependent disorder is closely related to the problem of magnetic
impurities in a superconductor\cite{abrikosov61}.

The present work is motivated by recent reports of measurements on trapped
fermionic gases in the presence of species-dependent 
gradients\cite{nichols17}.  
One could envision generating a similar situation in a condensed matter context
through the application of an in-plane
Zeeman-coupled magnetic field gradient.  The results here serve to
quantify the induced density gradient $\rho_\downarrow(x)$ and double
occupancy $D(x)$ for specific values of the interaction strength $U$,
temperature $T$, and chemical potential gradient $A_\uparrow$ which
would arise in such realizations.  Of
particular note are the following results: (i) The induced density
gradient in $\rho_\downarrow$ converges to that of $\rho_\uparrow$
when $U$ is made sufficiently large $U>U_*$.  (ii) Through a suitable
definition of the `characteristic density', defined below, data for
different gradients $A$ can be scaled together on a universal curve;
(iii) both the direct and induced density gradients initially grow
linearly with $A$, but while, at large enough $A$, the density
difference across the lattice approaches unity, if $U<U_*$ the induced
gradient reaches a plateau which does not reflect full
suppression/enhancement of density at the lattice edges; and
(iv) near neighbor spin correlations become ferromagnetic 
at the lattice edge for high gradients.

\section{Model and Methodology}

We consider the Hubbard Hamiltonian,
\begin{eqnarray}
H &= &-t \sum_{\langle ij \rangle \sigma}
\big( c^{\dagger}_{i\sigma} c^{\phantom{\dagger}}_{j\sigma}
+ c^{\dagger}_{j\sigma} c^{\phantom{\dagger}}_{i\sigma} \big)
\nonumber \\
&&+ U \sum_i \big( n_{i\uparrow} - \frac{1}{2} \big)
 \big( n_{i\downarrow} - \frac{1}{2} \big)
- \sum_{i\sigma} \mu_\sigma(i) \, n_{i\sigma},
\label{hubham}
\end{eqnarray}
with the usual kinetic energy, parameterized by $t$, describing
hopping between near neighbor sites $\langle i j \rangle$ on a 2D
square lattice, and on-site reuplsion $U$.  We choose $t=1$ to set the
scale of the energy.

The chemical $\mu$ is chosen to be spin-dependent:
\begin{eqnarray}
\mu_{\downarrow}(x)&=&0,
\nonumber \\
\mu_{\uparrow}(x)&=& A \big( x -\frac{n-1}{2} \big).
\label{chempot}
\end{eqnarray}
Here $x=0,1,2,\cdots ,n-1$ labels the $x$ coordinate of the site
$i=(x,y)$.  The form in Eq.~\ref{chempot} 
makes $\mu_\sigma = 0$ at the center of the
lattice and the average chemical potentials across the whole lattice,
$\langle \mu_\sigma \rangle = 0$.  Therefore, the filling is close to
$\rho_\sigma = 0.5$ (half-filling).  In most of our plots we have
chosen $n \times n$ lattices with $n=12$, and periodic boundary
conditions along $y$ and open ones along $x$.

It is simple to understand quantitatively the effect of the $\mu$
gradient using mean field theory (MFT) where the interaction term is
decoupled so that the down spin fermions see $+U \langle n_{i\uparrow}
\rangle$.  Since $\langle n_{i\uparrow}\rangle$ decreases as $x$
increases, due to $\mu_\uparrow$, down spin fermions see a lower
(more negative) chemical potential $\mu^{\rm mft}_{i\downarrow} = -U
\langle n_{x\uparrow} \rangle$ at one edge of the lattice, $i_x=0$,
than at the other, $i_x=n-1$.  This MFT-induced chemical potential
gradient introduces an effective density gradient for the down spin
fermions as well.

While one could formalize such a MFT treatment, we instead present here the
results of determinant Quantum Monte Carlo (DQMC) 
simulations,\cite{blankenbecler81}
which treat the effects of $U$ exactly.  In DQMC, the
interaction term is decoupled by introducing a space and imaginary
time dependent Hubbard-Stratonovich (HS) field.  The integration over
this field, with Monte Carlo, fully restores the effect of $U$, to
within the fully controlled errors introduced by the discretization of
the inverse temperature $\beta = L \Delta\tau$.  Here we choose
$\Delta\tau$ small enough so that these ``Trotter
errors"\cite{trotter59,suzuki76,fye86} are of the order of a few percent.
Most existing DQMC treatments of 
confined fermions\cite{paiva11}
or inhomogeneous systems generally\cite{paiva15},
do so within the local density approximation,
in order to analyze systems of larger size.
The spatial resolution of DQMC, however, allows
the exact treatment of gradients beyond the LDA.
We do so here.

Because of the nonvanishing $\mu_{\uparrow}(x)$, the weight associated
with individual HS field configurations can become negative leading to
a ``sign problem''\cite{loh90,troyer05}.  However, since we keep the
lattice roughly half-filled by the choice of $\mu_{\sigma}(i)$ in
Eq.~(\ref{chempot}), we find the sign problem to be quite mild.  A
similar mitigation of the sign problem is found in a bilayer system
when one sheet is electron doped and the other hole doped
symmetrically\cite{bouadim08}.

In the absence of gradients, powerful finite size scaling methods have
been developed both in classical and quantum Monte Carlo to extract
information about the thermodynamic limit from lattices of finite
size.  In the commonly studied case of confined lattices, the notions
of ``characteristic density'' and ``characteristic length'' were
developed \cite{rigol03,batrouni08,rigol09} in order to study and
compare systems with different sizes. The idea is to define
appropriate length scales associated with the nonuniformity in the
system and use them to define an effective characteristic density. In
Eqs.(\ref{hubham}) and (\ref{chempot}), the parameter $A/a$ has the
units of energy/length ($a$ is the lattice constant). Then a
characteristic length can be defined in the $x$-direction, $\xi \equiv
at/A$, while in the $y$-direction, which has no gradient, the
characteristic length remains $n$. The characteristic up spin density
is then $\tilde\rho_\uparrow \equiv N_\uparrow /(n\xi) = \rho_\uparrow
A n/t$, where $\rho_\uparrow \equiv N_\uparrow /n^2$. Similar
definitions apply for the down spins. We show below that systems with
different sizes will display the same properties if they are at a
common value of $\tilde\rho$, in analogy to what happens in traps
\cite{rigol03,batrouni08,rigol09}.

\section{Results- Local Density}

\begin{figure}[h!]
\centerline{\includegraphics[width=8cm]{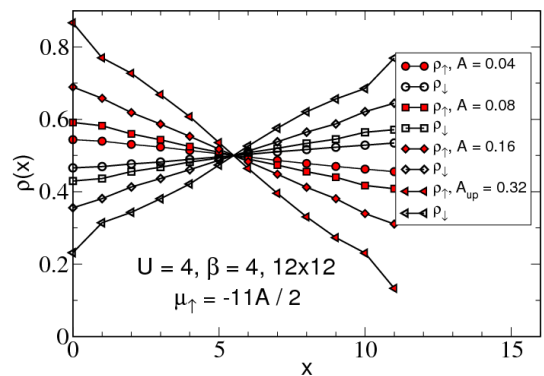}}
\caption{ Fermionic density gradients on a 12x12 lattice at $\beta=4$
  and $U=4$.  Only the up fermions experience a chemical potential
  gradient (see Eq.~1) but the interactions force the down fermion
  density to respond in an anti-correlated way.  }
\label{fig:rhovsxU4beta4}
\end{figure}

We begin by showing, in Fig.~\ref{fig:rhovsxU4beta4}, the density
gradients which arise from different up spin chemical potential
gradients $A$.  The interaction strength $U=4$ and inverse temperature
$\beta=4$ are kept fixed.  As suggested by the mean field theory
picture, the down spin fermions acquire a gradient, even though the
chemical potential they experience is spatially flat.  

\begin{figure}[h!]
  \centerline{\includegraphics[width=8cm]{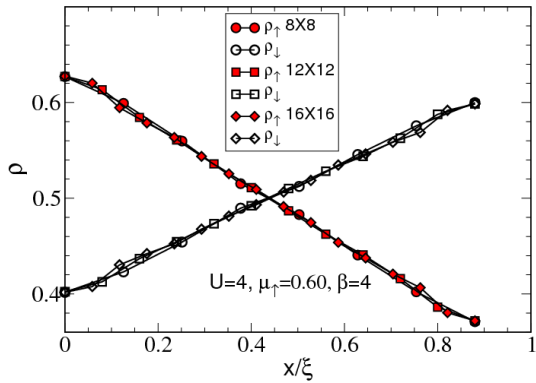}}
\caption{Density profile collapse for $n=8,12,16$ versus
    the characteristic coordinate, $x/\xi$. The simulations are at the
    same $\tilde\rho$ (see text).}
\label{fig:scaling}
\end{figure}

The data of Fig~\ref{fig:scaling} address the question of finite size
scaling by quantifying the density profiles for three system sizes
such that the characteristic density remains constant. We chose
$A=\delta\mu/(n-1)$ with $\delta\mu=0.88$, the chemical potential
difference between the edges of the system, kept constant leading to a
constant $\tilde\rho_\uparrow = \rho_\uparrow \delta\mu/t$. When the
density profiles are plotted against the characteristic position,
$x/\xi$, $\xi=at/A$, we obtain excellent collapse of the data.

\begin{figure}[t]
\centerline{\includegraphics[width=8cm]{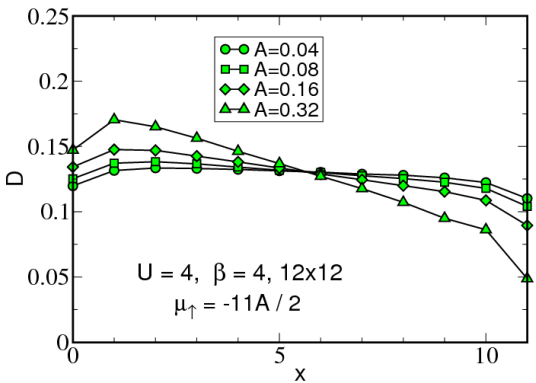}}
\caption{ Double occupancy $D$ versus position.  $U=4$ and $\beta=4$
  are fixed.  The up fermion chemical potential gradient, $A$, varies
  as indicated.  $D$ is fairly constant across the lattice for weak
  gradients, but is markedly asymmetric for stronger gradients, at
  this value of $U$.  }
\label{fig:dubvsxU4beta4}
\end{figure}

The double occupancy profile across the lattice is shown in
Fig.~\ref{fig:dubvsxU4beta4} for $U=4, \beta=4$ and different
gradients $A$.  This value of $U$ is not sufficient to impose a down
spin density gradient as large as the up spin gradient: The
enhancement of $\rho_\downarrow$ for $x=n-1$ is not developed fully
enough to compensate for the direct reduction in $\rho_\uparrow$ by
$A$, and hence $D$ is lower than at $x=0$.

\begin{figure}[t]
\centerline{\includegraphics[width=8cm]{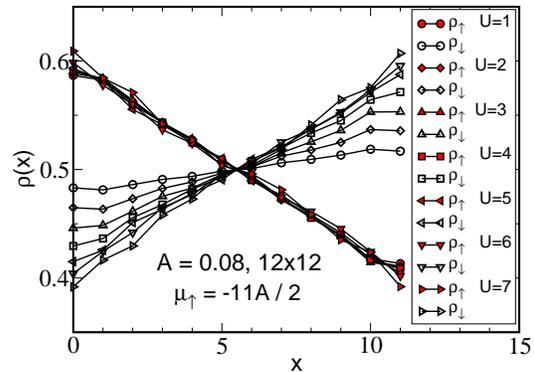}}
\caption{ Fermionic density gradients on a 12x12 lattice at $\beta=4$
  and $A=0.08$.  Only the up fermions experience a chemical potential
  gradient (see Eq.~1).  As is reasonable, when $U$ is small, the down
  gradient is less noticeable.  (Obviously it would vanish at $U=0$.)
  What is not completely expected is the almost complete insensitivity
  of $\rho_\uparrow(x)$ to $U$.  }
\label{fig:rhovsxA008beta4}
\end{figure}

\begin{figure}[t]
  \centerline{\includegraphics[width=8cm]{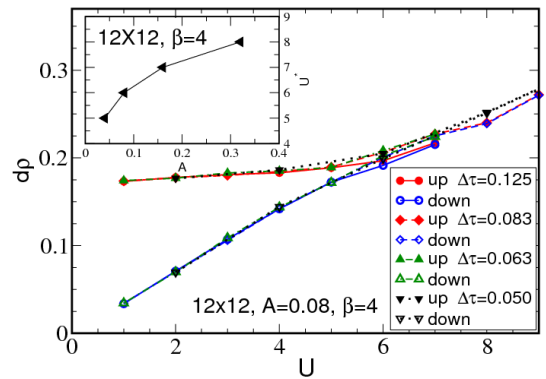}}
\caption{Absolute value of difference $d\rho$ between density at the
  two edges of box.  The induced difference $d\rho_\downarrow$
  achieves parity with the direct difference $d\rho_\uparrow$ at
  $U=U_*\sim 6$.  Data for different values of the inverse temperature
  discretization $\Delta \tau$ indicate that Trotter errors are, at
  largest, comparable to fluctuations associated with the statistical
  sampling.  The inset shows the dependence of $U_*$ on $A$.  }
\label{fig:drhovsUbeta4A008}
\end{figure}

Figure \ref{fig:rhovsxA008beta4} shows the effect of varying $U$ at
fixed up spin chemical potential gradient $A=0.08$ and inverse
temperature $\beta=4$.  The up spin density gradient is largely
independent of $U$ whereas, as expected, the induced down spin density
gradient develops more and more fully as $U$ grows.  This observation
is not completely captured within MFT.  One might expect that as a
nontrivial down spin profile develops, it would act to reinforce the
up spin density gradient by further lowering the energy cost for
putting up spin fermions at $x=0$ and raising it at $x=n-1$.  This is,
however, not a noticeable effect, at least for the parameters shown in
Fig.~\ref{fig:rhovsxA008beta4} (See also
Fig.~\ref{fig:drhovsUbeta4A008}.)

\begin{figure}[t]
  \centerline{\includegraphics[width=8cm]{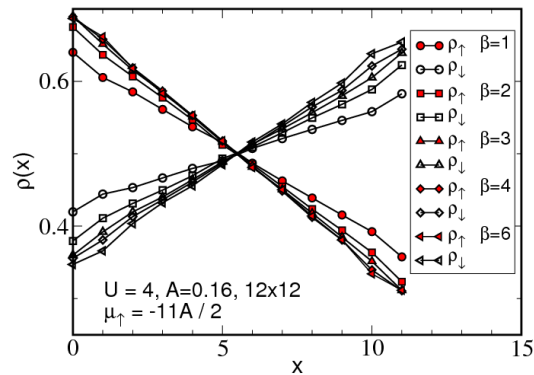}}
  \caption{Fermionic density gradients on a 12x12 lattice at $U=4$ and
    $A=0.16$ 
Only the up fermions experience a chemical
    potential gradient (see Eq.~1).  For this strength of chemical
    potential gradient, the density has achieved its low T value at
    $\beta \sim 4$.  }
\label{fig:rhovsxU4A016}
\end{figure}

Figure \ref{fig:rhovsxU4A016} quantifies the typical temperature
scales needed to establish the gradients for the $A$ and $U$ values in
Figs. \ref{fig:rhovsxU4beta4}-\ref{fig:rhovsxA008beta4}.  By the time
$\beta \sim 4$ (that is, $T=t/4$) the density gradients in the
presence of $U=4$ and $A=0.16$.

Collectively, Figs.~\ref{fig:rhovsxU4beta4},\ref{fig:rhovsxA008beta4}, and
\ref{fig:rhovsxU4A016} provide precise quantitative benchmarks for
the magnitude of interaction-induced gradients as
(i) the direct gradient; (ii) the interaction strength; and
(iii) the temperature, are varied, respectively.

\begin{figure}[t]
\centerline{\includegraphics[width=8cm]{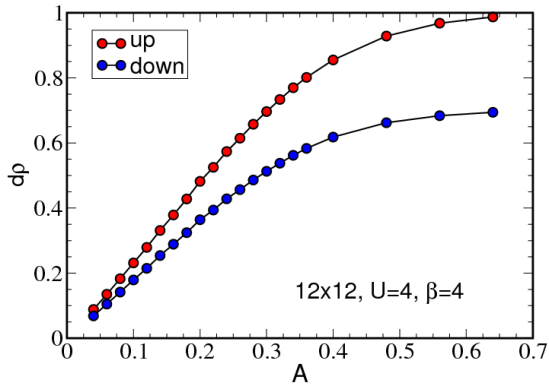}}
\caption{Absolute value of the difference $d\rho$ between the
  densities at the two edges of box.  At fixed $U$, as $A$ increases,
  the down spin gradient is less able to follow the up spin gradient,
  and ultimately saturates at less than the full value
  $d\rho_\uparrow=1$ corresponding to densities $\rho_\uparrow=0,1$ at
  the lattice edges.  Note that most of the preceding data were for
  the regime of relatively small $A$, where the slopes are linear.}
\label{fig:drhovsAbeta4U4}
\end{figure}

Inspection of the density gradients in the preceding figures already
suggest that they grow roughly linearly in $A$.  This is verified more
explicitly in Fig.~\ref{fig:drhovsAbeta4U4} by the data for the
overall edge-to-edge density differences
\begin{equation}
d\rho_\sigma = \big| \rho_\sigma(x=0) - \rho_\sigma(x=n-1) \big|
\end{equation}
A linear growth (albeit with a different slope for the two spin
species) occurs over a wide range of $A$.  This is followed, at large
$A$, by a plateau of the up spin density at full ``polarization",
where $\rho_\uparrow(x=0) \rightarrow 1$ and $\rho_\uparrow(x=n-1)
\rightarrow 0$, so that $d\rho_\uparrow \rightarrow 1$.  The down spin
polarization saturates at a reduced value.  Within the MFT picture
introduced earlier, the induced down spin chemical potential gradient
is bounded by $U$.  Since $U=4$ in Fig.~\ref{fig:drhovsAbeta4U4} is
only half the kinetic energy bandwidth, the induced gradient is
insufficient fully to localize the down spin fermions at the $x=n-1$
edge.  This accounts for the incomplete saturation of $d\rho_\downarrow$.

It is interesting to explore the variation of $d \rho_\sigma$ with
$U$, as done in Fig.~\ref{fig:drhovsUbeta4A008}.  One sees that the up
spin density gradient does in fact grow with $U$, as the MFT picture
discussed in the context of Fig.~\ref{fig:rhovsxA008beta4} might
suggest, but does so only weakly.  The induced down spin gradient
evolves much more rapidly, ultimately achieving, for $A=0.08$, parity
at $U = U_* \sim 6$, where the up and down density gradients fully
match up.  The inset shows the dependence of the value of $U$ at which
the up and down density gradients become equal, as a function of the
size of the gradient.  The dependence is seen to be rather weak.  We
believe this is so because $U_*$ is set not only by $A$, but also by
the quantum fluctuations (that is, the fermionic hopping parameter
$t$) which act to equalize the density across the lattice.

\section{Results- Local Spin Correlations}

In this section, the behavior of the local AF correlations
across the sample box is described.  Naively, as is the case for
fermions confined in a quadratic potential, one expects the
antiferromagnetic spin response to be maximal where the local density
is close to half-filling.  Here, with the choice of gradient,
Eq.~\ref{chempot}, this corresponds to the equipotential line, $\mu
=0$, in the center of the box.  This expectation is confirmed, and, in
addition, we discuss edge effects due to open boundary conditions in
the gradient ($\hat x$) direction.

We define the correlation function between a spin on site $(i_x,i_y)$
and one that is separated by a distance $j_y$ in the ($\hat y$)
direction, transverse to the gradient (i.e. at the same chemical
potential),
\begin{align}
c^{\alpha \alpha}(i_x,j_y) &=  
{\big \langle} \, 
 S^\alpha_{i_x, i_y+j_y}  \,
 S^\alpha_{i_x, i_y}  \,
{\big \rangle}
\nonumber \\
S^\alpha_{i_x,i_y} &= 
 \left(  \begin{array}{cc}
 c_{i_x i_y \uparrow}^{\dagger} &
 c_{i_x i_y \downarrow}^{\dagger} \\
 \end{array} \right) 
  \sigma^\alpha 
   \left(  \begin{array}{c}
  c_{i_x i_y \uparrow} \\
  c_{i_x i_y \downarrow} 
  \end{array} \right) \,\, ,
\label{eq:spincorrdef}
\end{align}
where $\sigma^\alpha$ is a Pauli spin matrix.  Because of periodic
boundary conditions in the $\hat y$-direction, this quantity is
independent of $i_y$, as suggested by the notation, which also serves
to emphasize the distinction between the two lattice directions, only
one of which is subject to the chemical potential gradient.  The local
moment is given by $\langle \, m^2(i_x) \, \rangle = c(i_x,j_y=0)$,
and the near neighbor spin correlator by $c(i_x,j_y=1)$.  In the data
which follow, we average $c^{\alpha\alpha}$ over the three
rotationally equivalent spin directions $\alpha=x,y,z$.

\begin{figure}[t]
  \centerline{\includegraphics[height=8.0cm,width=8.0cm]{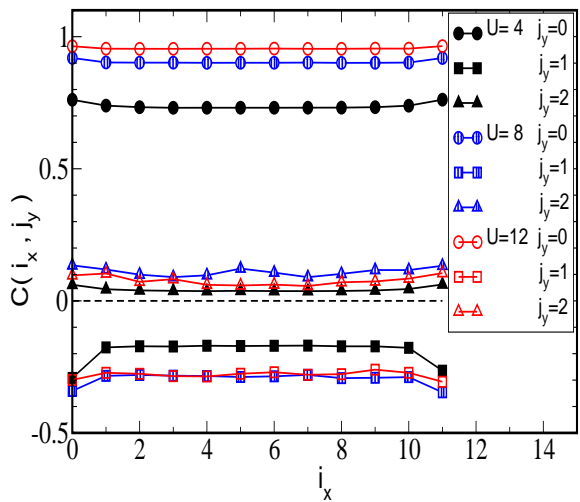}}
\caption{  Spin-spin correlation function $c(i_x,j_y)$
    at three values of the on-site interaction $U=4, 8, 12$.  Here
    $\beta=4$, the gradient $A_\uparrow=0$, and the lattice is $12
    \times 12$.  Magnetism is enhanced at the edges of the lattice
    $i_x=0$ and $i_x=n-1$ as a consequence of the OBC.  } 
\label{fig:cjyA00tweak}
\end{figure}

Figure \ref{fig:cjyA00tweak} exhibits the spin correlators for
$j_y=0,1,2$ as functions of position in the $\hat x$ direction in the
{\it absence} of a gradient.  A characteristic AF pattern is observed,
with a large, positive local moment ($j_y=0)$, negative near-neighbor
($j_y=1)$ correlations, and positive next near-neighbor ($j_y=2)$
correlations.  The absolute value of $c(i_x,j_y)$ decreases with $j_y$
since the temperature $T=t/4$ is not much below $J \sim t^2/U$.
Nevertheless, one can observe some of the crucial characteristic
features of AF correlations in the Hubbard Hamiltonian: a local moment
which increases monotonically with $U$, and a maximum in magnetism at
intermediate $U\sim 8$ reflecting the initial growth of magnetism at
weak $U$ followed by a decline at finite $T$ at strong coupling as the
exchange constant $J$ falls.

It is useful to understand this behavior of $c(i_x,i_y)$ at
$A_\uparrow=0$ because, even in the absence of a gradient,
translational invariance is still broken by the open boundary
conditions in the $\hat x$ direction.  This manifests itself as an
{\it enhancement} of magnetism at the boundaries ($i_x=0$ and
$i_x=n-1$) which is especially evident in the near-neighbor
correlation at $U=4$.  Such enhancement is known to occur in a variety
of materials, and can be ascribed either to the reduced coordination
at the surface and the associated lowering of quantum fluctuations (as
is the case here), or to the presence of modified exchange constants
due to surface relaxation or reconstruction\cite{zhang95}.

\begin{figure}[t]
  \centerline{\includegraphics[height=8.0cm,width=8.0cm]{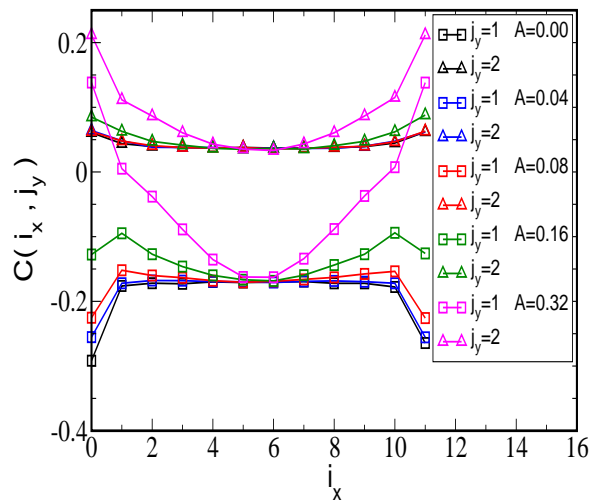}}
\caption{  Spin-spin correlation function $c(i_x,j_y)$
    at fixed $U=4$ and different values of the gradient $A=0.00, 0.04,
    0.08, 0.16$, and $0.32$.  Here $\beta=4$, and the lattice is $12
    \times 12$.    }
\label{fig:cjyU4}
\end{figure}

Having exhibited this surface enhancement, Fig.~\ref{fig:cjyU4} shows
the effect of different gradients on the spin correlations at $U=4$.
Consider first the near neighbor correlation $c(i_x,j_y=1)$, which
becomes less negative as $A_\uparrow$ increases: the gradient, in
combination with the on-site repulsion, polarizes the fermionic
density and suppressed antiferromagnetism.  Indeed, for sufficiently
large values of the gradient, the near-neighbor spin correlator
becomes {\it positive} (ferromagnetic) at the edges of the trap, while
remaining negative, with a value which is independent of $A_\uparrow$,
at the lattice center $i_x=(n-1)/2$.  (The constancy of
$c(i_x=(n-1)/2,i_y)$ can be viewed as an indication that the local
density approximation is valid: the magnetism depends only on the
density, which are $\rho_\sigma=1/2$, and not strongly on the
gradient.)  The next near-neighbor correlator, $c(i_x,j_y=2)$, which
is already ferromagnetic in the absence of a gradient, becomes larger
in magnitude with increasing $A_\uparrow$.

The position and spin-dependent chemical potentials $\mu_\sigma(x)$ in
Eq.~\ref{chempot} can equivalently be written as a spin-{\it
  independent} (but still spatially varying) chemical potential,
combined with a position dependent Zeeman field,
\begin{align}
\mu'(x) = B(x) = -\frac{A}{2} \big(x - \frac{n-1}{2} \big).
\label{zeemanrep}
\end{align}
Equation~\ref{zeemanrep} is easily seen to be equivalent to
Eq.~\ref{chempot} by noting that $\mu_\uparrow(x) = B(x)+\mu'(x)$ and
$\mu_\downarrow(x) = -B(x)+\mu'(x)$.  This provides an alternate
understanding of the spin polarization, and hence ferromagnetic
correlations at the lattice edges.

\section{Conclusions}
In this paper, we explored the effect of a spin-dependent chemical
potential gradient on a confined electron gas.  {\it Homogeneous}
spin-dependent potentials (Zeeman fields) have, of
course, been extensively studied within the context of pairing in
spin-imbalanced systems (Fulde-Ferrell-Larkin-Ovchinnikov
superconductivity\cite{fulde64,larkin64,wolak12}).  Similarly, as
noted in the introduction, spatially disordered spin-dependent
chemical potentials can produce novel superconductivity in the
presence of attractive interactions. They also have been shown to have
interesting consequences for metal-insulator transitions (MIT),
sharing the qualitative behavior of bond randomness, but behaving
quite distinctly from spin-independent disorder\cite{denteneer01}. The
combination of different types of spin-dependent and independent
fields likewise can drive systems across the MIT \cite{denteneer03}.

The anticorrelated gradient of charge of two spin species caused by an
on-site repulsion $U$ is similar in physical origin, and
qualitative consequences, to the transfer of charge between distinct
orbitals in condensed matter systems.  In a three-band model of
cuprate superconductors, for example, a difference in oxygen $p$ and
copper $d$ single particle energy $\Delta_{pd}=\epsilon_p -
\epsilon_d$ leads to an `orbital occupation gradient'.  An interesting
interplay then occurs between $\Delta_{pd}$ and the intersite Coulomb
interaction $V_{pd}$.  For example, as the total density of the system
increases, there can be a counterintuitive decrease in oxygen $p$
occupation\cite{scalettar91}. Thus, although the realization of
distinct potentials for different spin species is considerably more
challenging in the condensed matter context than in trapped atomic
systems, some of the underlying physics might be quite analogous in
considering the ``orbital label" rather than the spin one.

 Evaluation of local spin correlations, and their
  dependence on local density, has been a crucial objective of QMC
  studies of trapped fermionic gases, since this leads to an
  identification of the spatial extent of the region in which long
  range AF might be observed experimentally
  \cite{hart15,duarte15,cheuk16,mazurenko17}.  We have examined here
  the case of a linear gradient, complementing previous work in
  quadratic traps.  Most studies of a quadratic trap consider
  situations in which the density falls to zero at the box edge, and
  hence spin correlations are small.  A key observation in this work
  is that, in the case of moderate gradients in which the density
  remains nonzero at the box edge, there can be an enhancement of AF
  correlations due to the reduction of quantum fluctuations at the
  `surface' of the sample caused by the open boundary conditions. 
These correlations ultimately become ferromagnetic
at large gradients.

A particle-hole transformation $c^{\phantom{\dagger}}_{i\uparrow}
\rightarrow (-1)^i c^\dagger_{i\uparrow} $ maps the repulsive Hubbard
Hamiltonian onto the attractive Hubbard
Hamiltonian\cite{scalettar89,moreo07}.  Since $n_{i\uparrow}
\rightarrow 1-n_{i\uparrow}$, the chemical potential gradient reverses
sign, as does the associated up spin density gradient.  Since there is
no transformation of the down spin operators, the induced down spin
gradient is unchanged in sign, and we see that, as expected, in an
attractive model $d\rho_\downarrow/dx$ and $d\rho_\uparrow/dx$ have
the same sign.


\begin{acknowledgements}
  This work was supported by the Department of Energy under grant
  number DE-SC0014671.
\end{acknowledgements}

\end{document}